\documentclass[twocolumn,epjc3]{svjour3mod}
\usepackage{scrhack}
\usepackage{fix-cm}
\usepackage{amsmath,amssymb,amsfonts,dcolumn,mathcomp,slashed,dsfont}
\usepackage{graphicx}
\usepackage{bm}
\usepackage{xcolor}
\usepackage{csquotes}
\usepackage{units}
\usepackage{flushend}
\usepackage{cite}
\usepackage[colorlinks,citecolor=blue,urlcolor=blue,linkcolor=blue]{hyperref}
\usepackage{ragged2e}
\usepackage{comment}
\usepackage{subcaption}
\usepackage{booktabs}
\usepackage{physics}

\apptocmd\thebibliography\justifying{}{}

\newcommand{\A}{\mathcal{A}}
\newcommand{\B}{\mathcal{B}}

\newcommand{\F}{\mathcal{F}}
\newcommand{\G}{\mathcal{G}}

\renewcommand{\L}{\mathcal{L}}
\newcommand{\M}{\mathcal{M}}

\newcommand{\Order}{\mathcal{O}}
\newcommand{\disc}{{\rm disc}\,}
\renewcommand{\dd}{{\rm d}}

\newcommand{\GeV}{\,\text{GeV}}
\newcommand{\MeV}{\,\text{MeV}}
\newcommand{\fm}{\,\text{fm}}
\newcommand{\bsp}{\begin{sloppypar}}
\newcommand{\esp}{\end{sloppypar}}

\graphicspath{{./plots/}{./figures/}}

\begin{document}

\title{Polarizabilities from kaon Compton scattering}

\author{
Dominik Stamen\thanksref{Bonn,e1}
\and
Jan Luca Dammann\thanksref{Bonn}
\and
Yannis Korte\thanksref{Bonn}
\and
Bastian Kubis\thanksref{Bonn,e2}
}

\thankstext{e1}{stamen@hiskp.uni-bonn.de}
\thankstext{e2}{kubis@hiskp.uni-bonn.de}

\institute{Helmholtz-Institut f\"ur Strahlen- und Kernphysik (Theorie) and
   Bethe Center for Theoretical Physics,
   Universit\"at Bonn, 
   53115 Bonn, Germany \label{Bonn}
}

\date{}

\maketitle

\begin{abstract}
\bsp
The polarizabilities of light pseudoscalar mesons can be extracted from differential cross sections for Compton scattering near threshold.  While this has been accomplished for charged pions employing Primakoff reactions, a corresponding measurement for kaons will be affected by the presence of the $K^*(892)$ resonance not too far from threshold.  We propose a method to extend the energy range serviceable for this purpose by reconstructing the $K^*(892)$ contribution model-independently from its $K\pi$ intermediate state, using dispersion theory. We point out that, in contrast to the charged-pion analog, there is likely no strong hierarchy between sum and difference of electric and magnetic dipole polarizabilities; we discuss the sensitivity to disentangling both by improved experimental angular coverage. 
\esp
\end{abstract}

\section{Introduction}\label{sec:introduction}
In classical electrodynamics, electric and magnetic polarizabilities characterize the deformation response of a composite system in an external electromagnetic field. They appear as the coefficients of proportionality between fields and induced dipole moments~\cite{Holstein:2013kia}. In quantum field theory, the electric and magnetic polarizabilities, $\alpha$ and $\beta$, 
are defined from an expansion of the (Born-term-subtracted) Compton scattering amplitudes at threshold. Since there are no mesonic targets available, the Compton cross sections for mesons cannot be measured directly from photon scattering experiments. 

\bsp
Instead, in the charged-pion channel, high-energy pion--nucleus brems\-strahlung $\pi Z \to \pi Z \gamma$~\cite{Antipov:1982kz, Antipov:1984ez, COMPASS:2014eqi}, radiative pion photoproduction off the proton $\gamma p \to \gamma \pi^+ n$~\cite{Aibergenov:1986gi, Ahrens:2004mg}, and the crossed-channel two-photon reaction $\gamma\gamma \to \pi^+\pi^-$~\cite{PLUTO:1984efc, Courau:1986gn, Boyer:1990vu, CELLO:1992iai, Belle:2007ebm} have been measured. Historically, there have been tensions between the experimental extractions of the charged-pion polarizabilities and the theoretical predictions from chiral perturbation theory (ChPT)~\cite{Bijnens:1987dc,Bellucci:1994eb,Burgi:1996mm,Burgi:1996qi,Gasser:2005ud,Gasser:2006qa}, which have been extensively discussed in the literature~\cite{Moinester:2019sew}. The most recent measurement by the COMPASS experiment~\cite{COMPASS:2014eqi} is in perfect agreement with the ChPT prediction, and model dependencies for the extraction of pion polarizabilities may be sufficient to explain the tension with respect to earlier measurements. A dispersive calculation of the pion polarizabilities is in good agreement with the ChPT result~\cite{Pasquini:2008ep}; cf.\ also the new dispersive analysis in Ref.~\cite{Ermolina:2024daf}. Furthermore, new experiments are planned in order to clarify the current situation. For an extensive review, see Ref.~\cite{Moinester:2019sew}.
\esp

In principle, analogous reactions to all these can also be investigated for kaons. The photon--photon fusion processes $\gamma\gamma\to \bar{K}K$ have already been measured in the near-threshold region~\cite{ARGUS:1989ird,Belle:2013eck}.
The data however cannot realistically be used in order to extract polarizabilities, as the physical region is too far removed from the kinematical point that defines the polarizabilities to allow for a reliable expansion.  A promising setup for the extraction of the Compton scattering cross section for charged kaons from the Primakoff reaction~\cite{Primakoff:1951pj} is the proposed AMBER experiment located at CERN~\cite{Abbon:2014aex, Adams:2018pwt, Bernhard:2019jqz}. Using an intensified charged-kaon beam\footnote{The statistics using kaons is expected to be lower than in the pion case by a factor of 8~\cite{Friedrich2023}.} allows for a measurement of the $\gamma K^-\to \gamma K^-$ process. However, the different masses will render the extraction of polarizabilities more challenging than in the pion case: the definition of the polarizabilities amounts to a polynomial expansion at threshold, whose range of applicability is clearly limited, in particular by the lowest lying (vector) resonances.  As
${M_\rho}/{M_\pi} \approx 5.5$,
but
${M_{K^*}}/{M_K} \approx 1.8$,
a way to extend the range of applicability of the kaon Compton scattering amplitude to higher energies and thereby extract these quantities with increased accuracy is clearly desirable.  This is similar in spirit (although not in practical implementation) to the extraction of the chiral anomaly from Primakoff reactions~\cite{Hoferichter:2012pm,Hoferichter:2017ftn,Dax:2020dzg}, including data from the first resonance region.
In this article we propose such a method.

In order to dispersively reconstruct the $K^*$ resonance in the kaon Compton amplitude we use the solutions from the Primakoff reactions calculated in Ref.~\cite{Dax:2020dzg}. Therefore, with these two studies we in principle allow for a simultaneous data analysis of the charged-kaon polarizabilities, the chiral anomaly, and the radiative couplings of the $K^*$ resonance at AMBER.  

This article is structured as follows. In Sec.~\ref{sec:definitions}, we introduce the relevant definitions for the helicity amplitudes as well as their expansion in terms of dipole and higher-order polarizabilities. The Born terms and next-to-leading-order corrections in ChPT are discussed in Sec.~\ref{sec:Born}. Therein we also investigate a dispersive strategy to improve on the pion loops beyond the chiral expansion. In Sec.~\ref{sec:VMD}, effects of the vector $K^*(892)$ resonance are discussed: a vector-meson-dominance (VMD) model is used to estimate higher-order corrections to the polarizabilities; and in addition, $\gamma K\to K\pi$ amplitudes are used to reconstruct the $K^*$ resonance dispersively. Numerical results are presented in Sec.~\ref{sec:results}, and we propose a method to experimentally extract the kaon polarizabilities. A short discussion of neutral-kaon Compton scattering is provided in Sec.~\ref{sec:neutral_kaon}. In Sec.~\ref{sec:conclusions} we summarize our findings.

\section{Definitions}\label{sec:definitions}
The $S$-matrix element for the kaon Compton scattering process reads~\cite{Hoferichter:2011wk}
    \begin{align}
        \bra{\gamma(q_2,\lambda_2)K(p_2)}\ket{\gamma(q_1,\lambda_1)K(p_1)}&\notag\\
        =(2\pi)^4\delta^{(4)}(q_2+p_2-q_1&-p_1)\notag\\
        \cdot\bigg(\delta_{\lambda_1\lambda_2}+ie^2 \F_{\lambda_1\lambda_2}&(s,t)e^{i(\lambda_1-\lambda_2)\phi}\bigg)\,, \label{eq:matrix-element}
    \end{align}
with the two helicities $\lambda_1$ and $\lambda_2$ of the in- and outgoing photons, respectively. Furthermore, the azimuthal angle $\phi$ is explicitly separated. The Mandelstam variables read $s=(q_1+p_1)^2$, $t=(q_1-q_2)^2$, and $u=(q_1-p_2)^2$.
They fulfill the on-shell condition $s+t+u=2M_K^2$.
The Mandelstam variable $t$ can be expressed via $s$ and the cosine of the $s$-channel scattering angle $z_s=\cos\theta_s$ according to
    \begin{align}
        t(s,z_s)&=\frac{z_s-1}{2s}(s-M_K^2)^2\,.\label{eq:t-sz}
    \end{align}
We can separate the polarization vectors from the helicity amplitude $\F_{\lambda_1\lambda_2}$ via
    \begin{equation}
        \F_{\lambda_1\lambda_2}(s,t)=\epsilon_\mu(q_1,\lambda_1)\epsilon^*_\nu(q_2,\lambda_2)W^{\mu\nu}(s,t)\,,
    \end{equation}
and use the Bardeen--Tung--Tarrach procedure~\cite{Bardeen:1968ebo, Tarrach:1975tu} to define scalar amplitudes $\A(s,t)$ and $\B(s,t)$ without kinematic singularities. This results in~\cite{Garcia-Martin:2010kyn,Hoferichter:2011wk}
    \begin{align}
        W^{\mu\nu}(s,t)&=\A(s,t)\bigg(\frac{t}{2}g^{\mu\nu}+q_2^\mu q_1^\nu\bigg)\notag\\
        &\quad+\B(s,t)\bigg(2t\Delta^\mu \Delta^\nu - (s-u)^2g^{\mu\nu}\notag\\
        &\qquad\qquad\quad+2(s-u)(\Delta^\mu q_1^\nu + \Delta^\nu q_2^\mu)\bigg)\,,
    \end{align}
where $\Delta_\mu=(p_1+p_2)_\mu$. The amplitude is manifestly gauge and Lorentz invariant and fulfills the Ward identities $q_{1\mu} W^{\mu\nu}(s,t) = 0 = W^{\mu\nu}(s,t)q_{2\nu}$. Using the explicit form of the polarization vectors\footnote{Here we use the phase convention of Ref.~\cite{Edmonds:1960}.}
    \begin{align}
        \epsilon(q_1,\pm)=\mp\frac{1}{\sqrt{2}}\begin{pmatrix}
        0\\
        1\\
        \pm i\\
        0
        \end{pmatrix} ,
        ~~
        \epsilon(q_2,\pm)=\mp\frac{1}{\sqrt{2}}\begin{pmatrix}
        0\\
        \cos\theta_s\\
        \pm i\\
        -\sin\theta_s
        \end{pmatrix} ,
    \end{align}
and the four-momentum vectors
    \begin{align}
        q_1=\frac{s-M_K^2}{2\sqrt{s}}\begin{pmatrix}
        1\\
        0\\
        0\\
        1
        \end{pmatrix} ,
        ~~
        q_2=\frac{s-M_K^2}{2\sqrt{s}}\begin{pmatrix}
        1\\
        \sin\theta_s\\
        0\\
        \cos\theta_s
        \end{pmatrix} ,
    \end{align}
we obtain the helicity amplitudes in terms of the scalar ones by
    \begin{align}
        \F_{++}(s,t)=\F_{--}(s,t)&=4(M_K^4-su)\B(s,t) \,,\notag\\
        \F_{+-}(s,t)=\F_{-+}(s,t)&=-\frac{t}{2}\A(s,t)\notag\\
        &\quad +t(t-4M_K^2)\B(s,t)\,.\label{eq:helicity_amplitudes}
    \end{align}
One can also construct the amplitude in the center-of-mass system with Coloumb gauge, cf.\ Refs.~\cite{Kaiser:2008jm,Kaiser:2008ss}, according to\footnote{Note that the $T$-matrix element defined in Refs.~\cite{Kaiser:2008jm,Kaiser:2008ss} is related to $\F$ via $T=e^2 \F$.}
    \begin{align}
        \F(s,t)&=2 \Big\{-\bm{\epsilon}_1\cdot \bm{\epsilon}_2 \A^\text{KF}(s,t)\\
        &\quad+\bm{\epsilon}_1\cdot \mathbf{q}_2\, \bm{\epsilon}_2\cdot \mathbf{q}_1 \frac{2}{t}\big(\A^\text{KF}(s,t)+\B^\text{KF}(s,t)\big)\Big\} \,, \notag
    \end{align}
where the different scalar amplitudes are mapped onto each other by
    \begin{align}
        \A^\text{KF}(s,t)&=\frac{t}{4}\A(s,t)-\frac{(s-u)^2}{2}\B(s,t)\,,\notag\\
        \B^\text{KF}(s,t)&=2(s-M_K^2)^2\B(s,t)\,.
    \end{align}
The total differential cross section is given by
    \begin{equation}
        \frac{\dd \sigma}{\dd \Omega}=\frac{\alpha_\text{em}^2}{4s}\left(|\F_{++}(s,t)|^2+|\F_{+-}(s,t)|^2\right)\,,\label{eq:cross_section}
    \end{equation}
where $\alpha_\text{em}$ is the fine-structure constant.
This article uses the isospin limit with $M_K=0.496\GeV$ and $M_\pi=M_{\pi^\pm}$.\footnote{This convention was used in the pion--kaon scattering analysis of Refs.~\cite{Pelaez:2016tgi,Pelaez:2020gnd} and the $\gamma K\to K\pi$ analysis of Ref.~\cite{Dax:2020dzg}.}

Polarizabilities specify corrections to the Born terms.  Electric and magnetic dipole, quadrupole, and higher-order polarizabilities are defined as the expansion coefficients of the Born-term-subtracted amplitudes in powers of $t$ at fixed $s=M_K^2$~\cite{Guiasu:1978dz,Hoferichter:2011wk},
    \begin{align}
        \pm\frac{2\alpha_\text{em}}{M_K t}\widehat{\F}_{+\pm}(M_K^2,t)&=\left(\alpha_1\pm\beta_1\right)_{K^\pm} \notag\\ 
        &\quad+\frac{t}{12}\left(\alpha_2\pm\beta_2\right)_{K^\pm} +\order{t^2}, \notag\\
        \widehat{\F}_{+\pm}(s,t) &=\F_{+\pm}(s,t)-\F^\text{Born}_{+\pm}(s,t)\,.
        \label{eq:polar} 
    \end{align}
To illustrate the contributions of the polarizabilities that are most relevant in the forward or backward direction, respectively, one may define $z_\pm=1\pm z_s$ and expand the differential cross section to linear order in the polarizablities~\cite{Drechsel:1994kh,COMPASS:2014eqi}
    \begin{align}
    \label{eq:crosssection_direction}
        \left(\frac{\dd \sigma}{\dd \Omega}\right)_{P\gamma}&=\left(\frac{\dd \sigma}{\dd \Omega}\right)_{\text{Born}}-\frac{\alpha_\text{em}M_P^3(s-M_P^2)^2}{4s^2\left(sz_+ + M_P^2z_-\right)}\\
        &\hspace{.5cm}\cdot\left(z_-^2(\alpha_1-\beta_1)_P+z_+^2\frac{s^2}{M_P^4}(\alpha_1+\beta_1)_P\right)\,,\notag
    \end{align}
where $P$ is an arbitrary (charged) pseudoscalar meson.
\section{ChPT amplitudes and \texorpdfstring{$t$}{t}-channel cuts}\label{sec:Born}

At leading order in the chiral expansion, the ChPT amplitudes for charged-kaon Compton scattering reproduce the results of scalar quantum electrodynamics, hence they consist precisely of the Born terms only.  Nonvanishing contributions to the polarizabilities first appear at next-to-leading or one-loop order. 
Including these corrections, the scalar amplitudes read~\cite{Guerrero:1997rd}\footnote{Note the missing factor of 4 in Eq.~(2) of Ref.~\cite{Guerrero:1997rd} in comparison to Refs.~\cite{Bellucci:1994eb,Bijnens:1995vg,Bijnens:1996kk}. Additionally the loop correction has a wrong normalization by a factor of 2, cf.\ Appendix~B of Ref.~\cite{Lu:2020qeo}.}
    \begin{align}
        \A(s,t)&=\frac{1}{M_K^2-s}+\frac{1}{M_K^2-u}\notag\\
        &\quad+\frac{8}{F_K^2}\left(L_9+L_{10}\right) +\A^\text{loop}(t)\,,\notag\\
        \B(s,t)&=\frac{1}{2t}\left(\frac{1}{M_K^2-s}+\frac{1}{M_K^2-u}\right)\,,
    \end{align}
where
    \begin{align}
        \A^\text{loop}(t)&=- \frac{1}{8\pi^2F_K^2}\Bigg[\frac{3}{2}-\frac{2M_\pi^2}{t}\arctan^2 \bigg(\frac{1}{\sigma^\pi(t)}\bigg)\notag\\
        &\hspace{2cm}-\frac{4M_K^2}{t}\arctan^2\bigg(\frac{1}{\sigma^K(t)}\bigg)\Bigg]\,, \notag\\
        \sigma^P(t) &= \sqrt{\frac{4M_P^2}{t}-1} \,. \label{eq:loop}
    \end{align}
Using this result and Eq.~\eqref{eq:polar}, one can calculate the contribution to the polarizabilities, which yields~\cite{Donoghue:1989si,Guerrero:1997rd}
    \begin{align}
        (\alpha_1+\beta_1)^\text{ChPT}_{K^\pm}&=0 
        \,, \label{eq:ChPT-pols}\\    
        (\alpha_1-\beta_1)^\text{ChPT}_{K^\pm}&=\frac{8\alpha_\text{em}}{M_K F_K^2}(L_9+L_{10}) \notag \\
        &=\frac{M_\pi F_\pi^2}{M_K F_K^2} (\alpha_1-\beta_1)^\text{ChPT}_{\pi^\pm} + \Order(M_K) \notag\\
        &=1.1(2)\cdot 10^{-4}\fm^3 \Big\{1+\Order\big(M_K^2/\Lambda_\chi^2\big) \Big\}\,,\notag 
    \end{align}
where $\Lambda_\chi \approx 1\GeV$ is the chiral symmetry breaking scale.  
The result is in strict analogy to the pion polarizabilities, with the appropriate replacement of mass and decay constant, and hence can be expressed as a low-energy theorem in terms of the latter, up to $SU(3)$-breaking effects. 
Note that for this specific combination it is advantageous to measure the cross section in the backward direction, cf.\  Eq.~\eqref{eq:crosssection_direction}.
For the numerical estimate, we have employed the next-to-leading-order chiral prediction for the pion polarizabilities from Ref.~\cite{Gasser:2006qa}. The uncertainty therein is a chiral $SU(2)$ uncertainty, while the kaon polarizabilities are affected by (unknown) chiral $SU(3)$ corrections as indicated in Eq.~\eqref{eq:ChPT-pols}.
We will neglect the uncertainty on the pion polarizabilities in the following, and instead give estimates for the kaon ones.

The loop contributions in Eq.~\eqref{eq:loop} show nontrivial analytic structures: they include cuts due to two-pion and two-kaon intermediate states, with branch points at $t=4M_\pi^2$ and $t=4M_K^2$, respectively.  These obviously limit the range of convergence of the polynomial polarizability expansion around $t=0$, Eq.~\eqref{eq:polar}: a representation of the Compton amplitude (beyond Born terms) in terms of a polynomial is bound to fail beyond those thresholds.  
In the extraction of polarizabilities from charged-pion Compton scattering, the pionic one-loop corrections have therefore been retained~\cite{Kaiser:2008ss,COMPASS:2014eqi}, and the same is clearly advisable in the investigation of kaon Compton scattering: according to Eq.~\eqref{eq:t-sz}, the maximum value of $|t|$ probed at fixed $s$ is given by $(s-M_K^2)^2/s$, which reaches $4M_\pi^2$ already for $\sqrt{s} \approx 0.65\GeV$. 
In contrast, a polynomial expansion of the kaon loops, with their much higher threshold, is well justified.
We will discuss a further improvement of the $t$-channel cut in the following. 

\begin{figure}[t]
        \centering
        \includegraphics[width=0.2\textwidth]{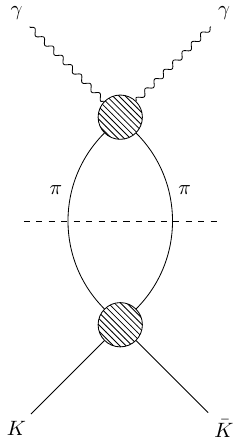}
        \caption{Dispersive reconstruction of the $t$-channel contribution to kaon Compton scattering, including the $\pi\pi$ intermediate states. The blobs denote the full $\gamma\gamma\to\pi\pi$ and $\pi\pi\to\bar{K}K$ $S$-wave amplitudes.}
        \label{fig:dispersive_t-channel}
    \end{figure}

The $t$-channel pion-loop contribution to kaon Compton scattering is of total isospin $I=0$, and, at next-to-leading order, a pure $S$-wave.  It is well known that the $I=0$ $S$-wave pion--pion system is particularly prone to sizeable rescattering corrections due to the proximity of the $f_0(500)$ resonance (cf., e.g., Ref.~\cite{Pelaez:2015qba} and references therein).  In order to control this effect reliably and prevent the nonanalyticities from perturbing the extraction of polarizability effects, we improve on the description of the pion--pion intermediate state by employing a dispersion-theoretical description of $\gamma\gamma\to\pi\pi\to K^+K^-$. This is shown diagrammatically in Fig.~\ref{fig:dispersive_t-channel}.

One may wonder why we concentrate on the lowest cut due to $\pi\pi$ intermediate states only, and do not resum all $t$-channel rescattering completely, given that the formalism is based on a coupled-channel treatment of $\gamma\gamma\to\pi\pi$ and $\gamma\gamma\to(\bar KK)_{I=0}$~\cite{Garcia-Martin:2010kyn} anyway.  As we are (mainly) interested in charged-kaon Compton scattering, the corresponding amplitude does not only consist of $t$-channel isospin $I=0$, but also contains $I=1$, which is the coupled-channel system of $\gamma\gamma\to\pi\eta$ and $\gamma\gamma\to(\bar KK)_{I=1}$~\cite{Danilkin:2017lyn,Lu:2020qeo,Schafer:2023qtl}. The reason is that both $\pi\eta$ and $\bar KK$ cuts are much further away from $t=0$, and do not affect the radius of convergence in a $t$-expansion for energies up to the $K^*(892)$ resonance.  Furthermore, the $\pi\eta$ singularity affects the amplitude far less, as $\gamma\gamma\to\pi\eta$ is famously suppressed in the chiral expansion~\cite{Ametller:1991dp,Jetter:1995js,Gan:2020aco}. Finally, including the kaon loops dispersively would lead to intricate dependencies, as their reconstruction depends on subtraction constants given by the very kaon polarizabilities we aim to extract, see Eq.~\eqref{eq:dispersive-pipi} below.

The $\gamma\gamma\to(\bar{K}K)_{I=0}$ $S$-wave amplitude $k^0$ contributes to
the helicity amplitudes according to~\cite{Garcia-Martin:2010kyn,Korte2024}
    \begin{align}
        \F_{+-}^{\pi\text{-disp}}(s,t)&=-\frac{t}{2}\A^{\pi\text{-disp}}(t)
        =\frac{1}{\sqrt{2}}k^0_{++}(t)\,,\notag\\
        \F_{++}^{\pi\text{-disp}}(s,t)&=0\,.
    \end{align}
Note that the helicity indices on the crossed-channel amplitudes denote the photon helicities in that channel. Furthermore, the factor $-1/\sqrt{2}$ is needed to relate the $I=0$ amplitude to the charge basis. 
The Born-term-subtracted coupled-channel amplitudes are given by~\cite{Garcia-Martin:2010kyn,Dai:2014zta,Korte2024}
    \begin{align}
        \label{eq:dispersive-pipi}
        \begin{pmatrix}h^0_{++}(t)\\k^0_{++}(t)\end{pmatrix}&=\mathbf{\Omega}(t)\Bigg\{\begin{pmatrix}a^\pi_t\\a^K_t\end{pmatrix}t\\
        &\hspace{-0.5cm}+\frac{t^2}{\pi}\Bigg(\sum_V \int_{-\infty}^{t_V}\dd x \frac{\mathbf{\Omega}^{-1}(x)}{x^2(x-t)}\Im\begin{pmatrix}h^{0,V}_{++}(x)\\k^{0,V}_{++}(x)\end{pmatrix}\notag\\
        & \quad- \int_{4M_\pi^2}^{\infty}\dd x \frac{\Im\mathbf{\Omega}^{-1}(x)}{x^2(x-t)}\begin{pmatrix}h^{0,\text{Born}}_{++}(x)\\k^{0,\text{Born}}_{++}(x)\end{pmatrix}\Bigg)\Bigg\}\,,\notag 
    \end{align}
where
    \begin{align}
      t_V=-\frac{1}{M_V^2}(M_V^2-M_\pi^2)^2\,,
    \end{align}
and $h^0$ denotes the $\gamma\gamma\to(\pi\pi)_{I=0}$ $S$-wave amplitudes. The vector-exchange and Born-term contributions can be found in the literature, cf.\ Ref.~\cite{Garcia-Martin:2010kyn}.
Thereby, suppressing the isospin index (since all elements have $I=0$), the Omnès matrix has the following components~\cite{Daub:2015xja,Ropertz:2018stk}
    \begin{align}
        \mathbf{\Omega}(t)=\begin{pmatrix}\Omega_{\pi\pi\to\pi\pi}(t)&\Omega_{\pi\pi\to\bar{K}K}(t)\\
        \Omega_{\bar{K}K\to\pi\pi}(t)&\Omega_{\bar{K}K\to\bar{K}K}(t)\end{pmatrix}\,.
    \end{align}
We now decouple and reduce it to only the $\pi\pi$ intermediate state explicitly, as shown diagrammatically in Fig.~\ref{fig:dispersive_t-channel}. Therefore, we set $a^K_t=k_{++}^{0,V}=k_{++}^{0,\text{Born}}=0$.
The system is now independent of the subtraction constant $a^K_t$, fixed from kaon polarizablities, and only depends on $a^\pi_t$, which is related to pion polarizabilities by
    \begin{equation}
        a^\pi_t=-\frac{M_\pi}{2\alpha_\text{em}} \frac{1}{\sqrt{3}}\left[2(\alpha_1-\beta_1)_{\pi^\pm} + (\alpha_1-\beta_1)_{\pi^0} \right]\,.
    \end{equation}

\begin{figure}[t]
        \centering
        \fontsize{12pt}{14pt} \selectfont
        \scalebox{0.65}{\input{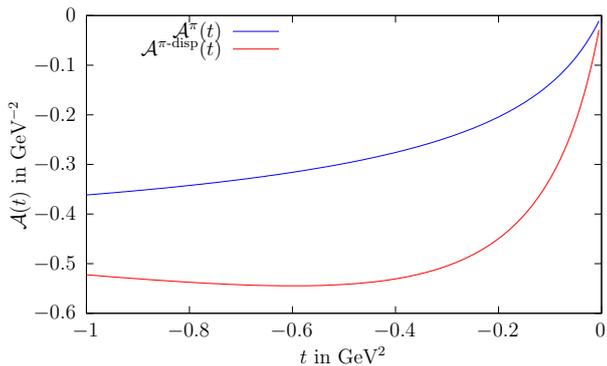}}
        \caption{Comparison of the pion loop in Eq.~\eqref{eq:pion_chpt} and the corresponding dispersive solution given by $\A^{\pi\text{-disp}}(t)$. The relevant physical range for $\sqrt{s}<0.85\GeV$ is $-0.31\GeV^2<t\leq 0\GeV^2$.}
        \label{fig:comparison_disp_chpt}
    \end{figure}

We now compare $\A^{\pi\text{-disp}}(t)$ to the ChPT result for the pion loops, 
    \begin{align}
        \A^\pi(t)&=-\frac{1}{8\pi^2F_K^2}\left(\frac{1}{2}-\frac{2M_\pi^2}{t}\arctan^2 \bigg(\frac{1}{\sigma^\pi(t)}\bigg)\right)\,,\label{eq:pion_chpt}
    \end{align}
in Fig.~\ref{fig:comparison_disp_chpt}. There the derivative at small $t$ can be related to the quadrupole polarizabilities via Eq.~\eqref{eq:polar}. It is known that the next-to-next-to-leading order ChPT result gives sizable corrections for the pion quadrupole polarizabilities~\cite{Gasser:2006qa}. As there is no calculation at that order for the kaons, we cannot make a quantitative comparison. However, corrections on the order shown in Fig.~\ref{fig:comparison_disp_chpt} are certainly not unexpected. The scalar function $\A$ including the dispersively reconstructed pion--pion intermediate state, but the $\bar KK$ one in the one-loop approximation, finally reads
    \begin{align}
        \A^\text{disp}(t)&=-\frac{\sqrt{2}}{t}k^0_{++}(t) \\
        &\quad-\frac{1}{8\pi^2F_K^2}\Bigg[1-\frac{4M_K^2}{t}\arctan^2\bigg(\frac{1}{\sigma^K(t)}\bigg)\Bigg]\,.\notag
    \end{align}
Note that the size of $\A^\text{disp}(t)$ does not play a major role for the extraction of the kaon polarizabilities presented in this paper, as the figures in Sec.~\ref{sec:results} look the same for both the ChPT and the dispersive solution.

\section{\texorpdfstring{$K^*$}{K*} resonance}\label{sec:VMD}
 
The effects of vector resonances (in the spirit of resonance saturation of chiral low-energy constants~\cite{Ecker:1988te}) 
only appear at next-to-next-to-leading-order ChPT~\cite{Bellucci:1994eb,Burgi:1996mm,Burgi:1996qi,Gasser:2005ud,Gasser:2006qa}. 
Assuming VMD and using an effective interaction Lagrangian for the vertex of the kaon, the photon, and the $K^*$~\cite{Garcia-Martin:2010kyn,Lu:2020qeo}
    \begin{equation}
      \L_\text{int}=e C_{K^*} \epsilon^{\mu\nu\alpha\beta} F_{\mu\nu} (\partial_\alpha K) K^*_\beta\,,
    \end{equation}
we find the helicity amplitudes~\cite{Lu:2020qeo} (cf.\ also Ref.~\cite{Bellucci:1994eb} for the analogous pion case)
    \begin{align}
        \F^V_{++}(s,t)&=C_{K^*}^2\left(M_K^4-su\right)\left[\frac{1}{M_{K^*}^2-s}+\frac{1}{M_{K^*}^2-u}\right],\notag\\
        \F^V_{+-}(s,t)&=C_{K^*}^2t\left[\frac{s}{M_{K^*}^2-s}+\frac{u}{M_{K^*}^2-u}\right]\,.\label{eq:vmd_amplitude}
    \end{align}
The coupling $C_{K^*}$ can be fixed with the $K^*\to K\gamma$ width using the values from the Review of Particle Physics~\cite{ParticleDataGroup:2022pth} via
    \begin{equation}
        \Gamma_{K^*\to K\gamma}=\alpha_\text{em} \frac{C_{K^*}^2}{6} \bigg(\frac{M_{K^*}^2-M_{K}^2}{M_{K^*}}\bigg)^3\,.
    \end{equation}
An expansion of Eq.~\eqref{eq:vmd_amplitude} according to Eq.~\eqref{eq:polar} leads to the magnetic polarizability 
    \begin{align}
        (\beta_1)^\text{VMD}_{K^\pm}&=\frac{4\alpha_\text{em}C_{K^*}^2M_K}{M_{K^*}^2-M_K^2}
        = \frac{24 M_{K^*}^3 M_K }{(M_{K^*}^2-M_K^2)^4} \Gamma_{K^*\to K\gamma}
        \notag\\
        &=0.36\cdot 10^{-4}\fm^3\,,\label{eq:VMD_pola}
    \end{align}
while no contribution to the electric one $(\alpha_1)^\text{VMD}_{K^\pm}$ is found.
This is in accordance with the fact that a pseudoscalar-to-vector transition requires a spin flip in the quark model, and hence a magnetic photon coupling.
Quantitatively, we observe that the vector-exchange correction reduces the $\mathcal{O}(p^4)$ ChPT result for $(\alpha_1-\beta_1)_{K^\pm}$, cf.\ Eq.~\eqref{eq:ChPT-pols}, by roughly 30\%.
In addition, the hierarchy $(\alpha_1+\beta_1) \ll (\alpha_1-\beta_1)$, still valid at full $\mathcal{O}(p^6)$ in ChPT for charged pions~\cite{Gasser:2006qa}, is modified very significantly: the ratio $(\alpha_1+\beta_1)/(\alpha_1-\beta_1)$ is increased to almost 50\%, while it remains at the level of around 3\% for the pions. 
As a result, the $(\alpha_1+\beta_1)_{K^\pm}$ contribution can in general not be neglected for kaon Compton scattering.

As the detailed studies of charged-pion polarizabilities at $\mathcal{O}(p^6)$ ChPT demonstrate~\cite{Bellucci:1994eb,Burgi:1996qi,Gasser:2006qa}, vector-exchange contributions alone only yield a fair estimate of the order of magnitude of such effects, but not a reliable quantitative assessment.   From the point of view of resonance saturation, axial-vector exchanges reduce the vector effects on $(\alpha_1-\beta_1)$, but add to them for $(\alpha_1+\beta_1)$, as they only contribute to $\alpha_1$ via
\begin{equation}
    (\alpha_1)_{K^\pm}^\text{axial}=\frac{4\alpha_\text{em}C_{K_1}^2M_K}{M_{K_1}^2-M_K^2} = \frac{24 M_{K_1}^3 M_K }{(M_{K_1}^2-M_K^2)^4} \Gamma_{K_1\to K\gamma}\,,
\end{equation}
where the Lagrangian for axial-vector to pseudoscalar and photon transitions reads~\cite{Garcia-Martin:2010kyn}
\begin{equation}
    \L_\text{axial} = 2 e C_{K_1} F_{\mu\nu} \partial^\mu K_1^\nu K\,.
\end{equation}
The main difficulty is to access the radiative charged axial-vector coupling, since there is no experimental data available. Using nonet symmetry one can relate the strange couplings to the non-strange ones via~\cite{Roca:2003uk,Garcia-Martin:2010kyn}
\begin{equation}
    C^2_{K_1^+(1270)}+C^2_{K_1^+(1400)}=C^2_{b_1^+(1235)}+C^2_{a_1^+(1260)}\,,
    \label{eq:axial_vector_symmetry}
\end{equation}
where one further needs to assume the Lipkin mechanism~\cite{Lipkin:1977uy} in order to extract the physical $K_1^+(1270)$ coupling. Moreover, the branching ratio for the $a_1^+(1260)$ is disputed in the literature~\cite{Zielinski:1984au,Condo:1993xa,CLAS:2008zko,COMPASS:2014qte}, with no precise modern experimental value available. A similar symmetry relation to Eq.~\eqref{eq:axial_vector_symmetry} can be tested for the radiative neutral strange axial-vector couplings,
\begin{equation}
    C^2_{K_1^0(1270)}+C^2_{K_1^0(1400)}=4C^2_{b_1^+(1235)}\,,
\end{equation}
which are overestimated by more than a factor of 2. Hence, we do not attempt to quantify the axial-vector contributions here.
Furthermore, loop corrections are sizable, as are the uncertainties due to variations of the scale at which resonance saturation is assumed to hold.  We take this into account below by associating a 50\% uncertainty with the VMD estimates for the polarizabilities, which is probably not even overly conservative.
The main difference between the kaon and pion cases is the size of the respective masses: higher orders in the polarizabilities are suppressed only by $M_K^2$ instead of $M_\pi^2$, making the former more susceptible to corrections by a factor of about 12.

    \begin{figure}[t]
        \centering
        \includegraphics[width=0.45\textwidth]{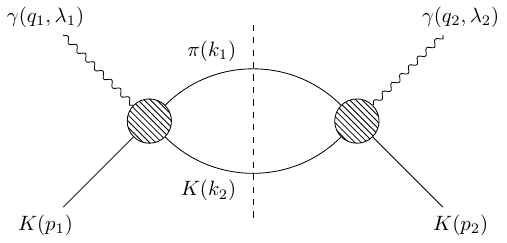}
        \caption{Compton scattering of kaons via an $K\pi$ intermediate state. The two photons have momenta $q_i$ and helicity $\lambda_i$, while the in- and outgoing kaon momenta are labeled by $p_i$. The internal momenta are $k_1$ for the pion and $k_2$ for the kaon.}
        \label{fig:dispersive_s-channel}
    \end{figure}

Apart from corrections to the polarizabilities, however, the appearance of resonances in the Compton amplitude limits the energy range that can sensibly be employed for the experimental extraction thereof.  For this specific problem, there is still a clear hierarchy in mass scales between the $K^*(892)$ and the first axial-vector state, the $K_1(1270)$.
We will therefore reconstruct the $K^*$ contribution dispersively via its two possible $K\pi$ intermediate states.

We show the exact form of the partial-wave expansion of the Compton helicity amplitudes in \ref{app:disc}.  There, we also derive 
the imaginary parts of the \mbox{$P$-waves} due to $K\pi$ intermediate states, see Fig.~\ref{fig:dispersive_s-channel}, for the $+\pm$ helicities, which read
    \begin{align}
        \Im\big(f_{1,\pm}&(s)\big)\notag\\
        =  \pm &\frac{1}{8\pi e^2}\frac{\lambda^{3/2}(s,M_\pi^2,M_K^2)}{72s^2}\notag\\
        &\cdot\left(|g_1^c(s)|^2+|g_1^n(s)|^2\right)\theta\big(s-(M_K+M_\pi)^2\big)\,,
    \end{align}
where $\lambda(a,b,c)=a^2+b^2+c^2-2(ab+bc+ac)$ is the Källén function.\footnote{Note that we include the factor of $e^2$ due to the amplitude definition in Ref.~\cite{Dax:2020dzg}.} Here $g_1^c$ and $g_1^n$ are the $\gamma K \to K \pi$ amplitudes with the charged and neutral kaons in the final state, respectively.\footnote{This form holds for both charged- and neutral-kaon Compton scattering, with $g_1^{c,n}$ replaced accordingly.}
Using an $n$-times subtracted dispersion integral we obtain the $\gamma K \to \gamma K$ $P$-wave amplitude
    \begin{align}
        f_{1,\pm}(s)&=P_{n-1}(s)\\
        &\quad+\frac{\left(s-M_K^2\right)^n}{\pi}\int_{M_K^2}^\infty\dd s' \frac{\Im(f_{1,\pm}(s'))}{\left(s'-M_K^2\right)^n(s'-s)}\,,\notag
    \end{align}
where we have subtracted at $s=M_K^2$.
We use the twice subtracted result from Ref.~\cite{Dax:2020dzg} for the $\gamma K\to K\pi$ amplitudes, which approach a constant for $s\to\infty$. 
Since we only need $g^i_1(s)$ in the low-energy region close to the $K^*$ resonance and are not interested in the high-energy behavior of this amplitude, we set $g^i_1(s)\to 1/s$ for large $s$, starting at $1\GeV$. With one subtraction the partial-wave amplitude then reads
    \begin{equation}
      f^1_{1,\pm}(s)=a^\pm_1\pm\bar{f}^1_{1}(s)\,,\label{eq:one_sub_disp}
    \end{equation}
where
    \begin{equation}
      \bar{f}^1_{1}(s)=\frac{(s-M_K^2)}{\pi}\int_{M_K^2}^{\infty}\dd s' \frac{\Im(f_{1,+}(s'))}{(s'-M_K^2)(s'-s)}\,.\label{eq:disp_integral}
    \end{equation}
The subtraction can be fixed via $a^\pm_1=f^1_{1,\pm}(M_K^2)$.

Reinserting the $P$-waves into the partial-wave decomposition for the helicity amplitudes leads to the dispersively reconstructed $K^*$ contributions
    \begin{align}
        \F_{++}^{K^*}(s,t)&=\frac{3}{4}\left(M_K^4-su\right) \left[f^1_{1,+}(s)+f^1_{1,+}(u)\right]\,,\notag \\
        \F_{+-}^{K^*}(s,t)&=-\frac{3}{4} t \left[ s f^1_{1,-}(s)+ u f^1_{1,-}(u)\right]\,.\label{eq:disp_amplitudes}
    \end{align}
The $u$-channel pieces are added by hand in order to render the amplitude crossing symmetric. 
Comparing to Eq.~\eqref{eq:vmd_amplitude}, we note that the VMD model can be brought into exactly the same form.
Applying Eq.~\eqref{eq:polar} then allows us to identify the connection between the subtraction constants and the polarizabilities extracted from the VMD model,
    \begin{equation}
        a_1^\pm =\frac{1}{3M_K\alpha_\text{em}}\left(\alpha_1\pm\beta_1\right)^\text{VMD}_{K^\pm}=\pm\frac{4}{3}\frac{C_{K^*}^2}{M_{K^*}^2-M_K^2}\,.\label{eq:subtractions}
    \end{equation}

To understand to what extent these subtraction constants can be replaced by the phenomenological polarizabilties that, ultimately, ought to serve as free parameters to be extracted from experiment, we need to discuss the structure of
our complete amplitude representation, which is of the following form.  In addition to the Born terms, we retain
\begin{enumerate}
    \item the dipole polarizabilities;
    \item dispersive representations of the leading singularities in $s$-, $t$-, and $u$-channels, which comprises the $K\pi$ $P$-waves (or the $K^*(892)$) in $s$- and $u$-channels, as well as the $\pi\pi$ $S$-wave (or the $f_0(500)$) in the $t$-channel;
    \item and the perturbative $t$-channel kaon loop to account for the missing piece in the above to reproduce the complete next-to-leading-order chiral representation.
\end{enumerate}
The Born-term-subtracted helicity amplitudes are therefore of the form
    \begin{align}
        \widehat{\F}_{+-}(s,t)&=-\frac{t}{2}\bigg[\frac{M_K}{\alpha_\text{em}}(\alpha_1-\beta_1)_{K^\pm}+\A^\text{disp}(t) \notag\\
        &\hspace{1cm}- \frac{3}{2}\left( t a_1^- + s \bar{f}^1_{1}(s) + u \bar{f}^1_{1}(u)\right)\bigg]\,,\notag\\
        \widehat{\F}_{++}(s,t)&=\frac{3}{4}(M_K^4-su) \bigg[\frac{2}{3M_K\alpha_\text{em}}(\alpha_1+\beta_1)_{K^\pm}\notag\\
        &\hspace{2.7cm}+\bar{f}^1_{1}(s) + \bar{f}^1_{1}(u)\bigg]\,.\label{eq:born_subtracted}
    \end{align}
Note that the dipole polarizabilities $(\alpha_1\pm\beta_1)_{K^\pm}$ now subsume the contributions induced by the subtraction constants of the dispersive $P$-wave. For the $+-$ helicity amplitude, however, a term proportional to $t^2 a_1^-$ is explicitly kept, since it induces a contribution to the quadrupole polarizabilities. These only come from the $K^*$ resonance. Here, $a_1^-$ is fixed by Eq.~\eqref{eq:subtractions}. Furthermore, in the $++$ helicity we keep the kinematic structure in front of the scalar amplitude, cf.\ Eq.~\eqref{eq:helicity_amplitudes}.

\section{Results}\label{sec:results}

\begin{figure}[t]
        \centering
        \fontsize{12pt}{14pt} \selectfont
        \scalebox{0.662}{\input{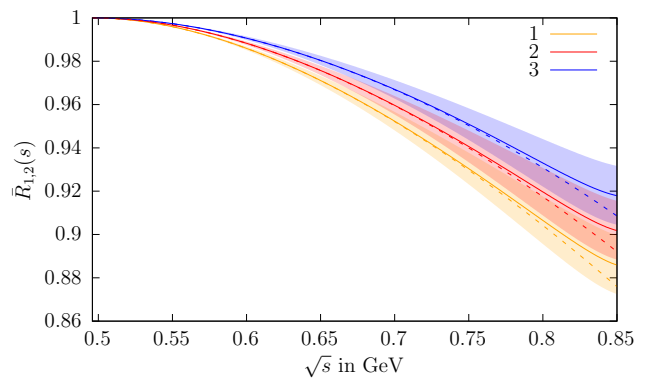}}
        \caption{Energy dependence of the integrated ratio of the differential cross sections over the backward half angle $\bar{R}_1(s)$ as defined in Eq.~\eqref{eq:integration_back}. The colors indicate the scenarios 1 to 3, which are described in the main text.
        The dashed lines represent the same input for the kaon polarizabilities as indicated by their color, but for the unphysical integrated ratio $\bar{R}_2(s)$.}
        \label{fig:integrate_back}
    \end{figure}

In analogy to Ref.~\cite{COMPASS:2014eqi}, we now define ratios of differential cross sections to extract the kaon polarizabilities and test the sensitivity to the latter. 
Here, the experimentally measured quantity is normalized and compared to a theoretical one.
We add the Born terms and calculate the differential cross sections via Eq.~\eqref{eq:cross_section}.
The ratio optimized for sensitivity to polarizability effects, retaining all other parts of the amplitude, then reads 
    \begin{align}
        R_1(s,t)&=\frac{\dd\sigma/\dd\Omega(s,t)}{\dd\sigma/\dd\Omega(s,t)\Big|_{\alpha_1=\beta_1=0}}
        \label{eq:ratio}\,,
    \end{align}
where the label indicates that $\alpha_1$ and $\beta_1$ are set to zero in Eq.~\eqref{eq:born_subtracted}, while all other terms are kept. For an analysis focusing on the extraction of the kaon polarizability \textit{difference}, this ratio can be integrated over backward angles, leading to
    \begin{align}
      \bar{R}_1(s)=\int_{-1}^{z_\text{cut}}\dd z_s R_1(s,t(s,z_s)) \label{eq:integration_back}\,.
    \end{align}
This backward-integrated ratio $\bar{R}_1(s)$, with cutoff angle $z_\text{cut}=0$, is shown in Fig.~\ref{fig:integrate_back}.
For the difference of kaon polarizabilities $(\alpha_1-\beta_1)_{K^\pm}$ we use three different scenarios.
We employ the ChPT prediction, Eq.~\eqref{eq:ChPT-pols}, and add $0.5$, $1$, and $1.5$ times the VMD induced corrections, Eq.~\eqref{eq:VMD_pola}, labeled by scenario 1 to 3, respectively.\footnote{Note that due to Eq.~\eqref{eq:VMD_pola} the two linear combinations have a different sign for the VMD contribution to the polarizabilities. Therefore, adding 1.5 times the VMD result leads to a smaller polarizability for $(\alpha_1-\beta_1)_{K^\pm}$ than 0.5 times the VMD result.} This corresponds to varying $(\alpha_1-\beta_1)_{K^\pm}$ by approximately $\pm20\%$ around the central value, given by the ChPT result with higher-order corrections estimated by VMD; in other words, the different full lines in Fig.~\ref{fig:integrate_back} indicate the required accuracy to extract $(\alpha_1-\beta_1)_{K^\pm}$ at the 20\% level.

On the other hand, the colored bands around the three central full curves depict a variation in \mbox{$(\alpha_1+\beta_1)_{K^\pm}$} from $0.5$ to $1.5$ times the VMD result.
While we will discuss strategies to actually disentangle $(\alpha_1\pm\beta_1)_{K^\pm}$ below, in the context of the extraction of the polarizability difference in backward directions, we regard the sum merely as a source of uncertainty.
We observe that these bands grow wider with increasing energy, start to overlap significantly around $\sqrt{s} \approx 700\MeV$, and around $800\MeV$ are so wide that the 50\% variation in $(\alpha_1+\beta_1)_{K^\pm}$ roughly makes up for the 20\% variation in $(\alpha_1-\beta_1)_{K^\pm}$: the impact of the \textit{sum} of polarizabilities on the extraction of the \textit{difference} is by far not negligible.  This is in notable contrast to the charged-pion case.  This effect can be reduced by choosing the angle more narrowly ($z_\text{cut}<0$), which will on the other hand limit the statistics. 
    \begin{figure}[t]
         \centering
        \fontsize{12pt}{14pt} \selectfont
        \scalebox{0.662}{\input{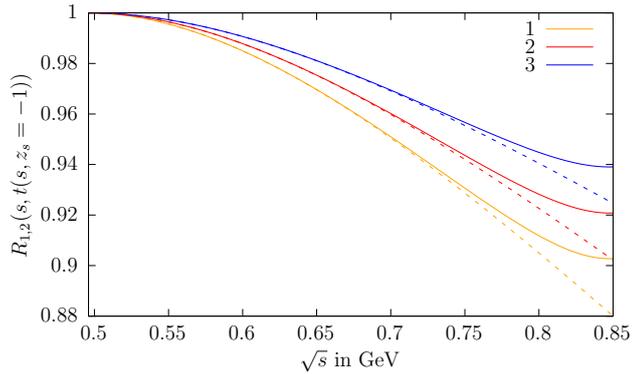}}
        \caption{Energy dependence of the ratio of the differential cross sections, Eq.~\eqref{eq:ratio}, for $z_s=-1$. Line styles as in Fig.~\ref{fig:integrate_back}.}
        \label{fig:zm1}
    \end{figure}
The extreme case is illustrated in Fig.~\ref{fig:zm1}: in strict backward direction,
for fixed $z_s=-1$, the amplitude is independent of the linear combination $(\alpha_1+\beta_1)_{K^\pm}$ and therefore  does not show any bands. However, it is unrealistic that an experiment will gather enough statistics for such a fixed angle.
Therefore, it will become crucial in the experimental analysis to optimize the interplay of both these quantities. 

We can define different ratios of differential cross sections to illustrate the necessity to retain the $K^*$ resonance in the polarizability extraction; all of them are similarly studied in forms integrated over backward angles. 
We define 
    \begin{align}
        R_2(s,t)&=\frac{\dd\sigma/\dd\Omega(s,t)\Big|_{K^*=0}}{\dd\sigma/\dd\Omega(s,t)\Big|_{\alpha_1=\beta_1=K^*=0}}\,, \notag\\
        \bar{R}_2(s)&=\int_{-1}^{z_\text{cut}}\dd z_s R_2(s,t(s,z_s))\,,\label{eq:R2}
    \end{align}
where $K^*=0$ denotes that we neglect the $K^*$ resonance and therefore do not include $\bar{f}_1^1$ and $a_1^-$ in Eq.~\eqref{eq:born_subtracted}. 
This is a purely theoretical or unphysical ratio (as the $K^*$ is omitted also in the numerator, which is impossible to measure experimentally), but it serves to illustrate the analogy to the pion case, where the $\rho$ resonance is outside of the relevant energy region:
its influence is suppressed because the energy range is chosen to be $\sqrt{s} < 3.5M_\pi$~\cite{COMPASS:2014eqi}, where the $\rho$ is far enough away ($M_\rho \approx 5.5 M_\pi$, or $3.5M_\pi \approx M_\rho - 2 \Gamma_\rho$). However, this still leads to an available energy range above threshold of approximately $350\MeV$.
$\bar{R}_2(s)$ hence shows the sensitivity of a charged-kaon polarizability extraction in a world where no $s$-channel resonance disturbs the range of validity of the expansion.

This ratio is also included for comparison in Fig.~\ref{fig:integrate_back} for the same three different values of $(\alpha_1-\beta_1)_{K^\pm}$ (but neglecting the variation in $(\alpha_1+\beta_1)_{K^\pm}$, which leads to extremely similar bands as for $\bar{R}_1(s)$), denoted by dashed lines.  We find that full and dashed lines, corresponding to $\bar{R}_1(s)$ and $\bar{R}_2(s)$, are extremely close up to roughly $\sqrt{s} \approx 800\MeV$: our effort to stabilize the ratio by inclusion of the $K^*$ is entirely successful, and data for higher energies can be made available for the polarizability extraction that way.  
Note that the energy range displayed in Figs.~\ref{fig:integrate_back}--\ref{fig:integrate_back_nodisp} corresponds rather precisely to the $350\MeV$ used for the pion polarizability extraction~\cite{COMPASS:2014eqi}.

Finally, we investigate the ratio
    \begin{align}
        R_3(s,t)&=\frac{\dd\sigma/\dd\Omega(s,t)}{\dd\sigma/\dd\Omega(s,t)\Big|_{\alpha_1=\beta_1=K^*=0}}\notag\,,\\
        \bar{R}_3(s)&=\int_{-1}^{z_\text{cut}}\dd z_s R_3(s,t(s,z_s))\,, \label{eq:R3}
    \end{align}
where the full, experimentally accessible, differential cross section is employed in the numerator, but normalized to a denominator omitting the $K^*$ effects.  This corresponds to the attempt to extract polarizabilities from real data, but ignoring the $K^*$ in the normalization expression.
We obviously expect $R_3$ to show strong deviations from unity when increasing $s$ towards the $K^*$ resonance energy, which are indeed visible in Fig.~\ref{fig:integrate_back_nodisp}: 
    \begin{figure}[t]
        \centering
        \fontsize{12pt}{14pt} \selectfont
        \scalebox{0.662}{\input{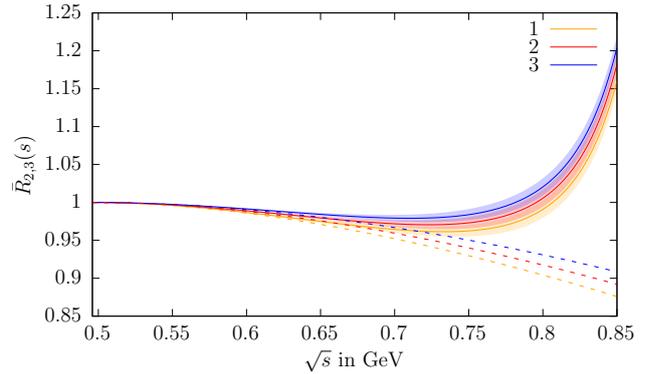}}
        \caption{Energy dependence of the integrated ratio of the differential cross sections over the backward half angle $\bar{R}_3(s)$ as defined in Eq.~\eqref{eq:R3}, neglecting the dispersively reconstructed $K^*$ in the denominator. 
        Line styles as in Fig.~\ref{fig:integrate_back}.}
        \label{fig:integrate_back_nodisp}
    \end{figure}
the onset of the resonant rise in $\bar{R}_3(s)$ obscures the polarizability sensitivity at the 20\% level already for energies as low as $\sqrt{s} \approx 650\MeV$, and quickly deteriorates even further above.
We therefore strongly suggest to use $\bar{R}_1$ for the extraction of kaon polarizabilities.

\begin{figure}[t]
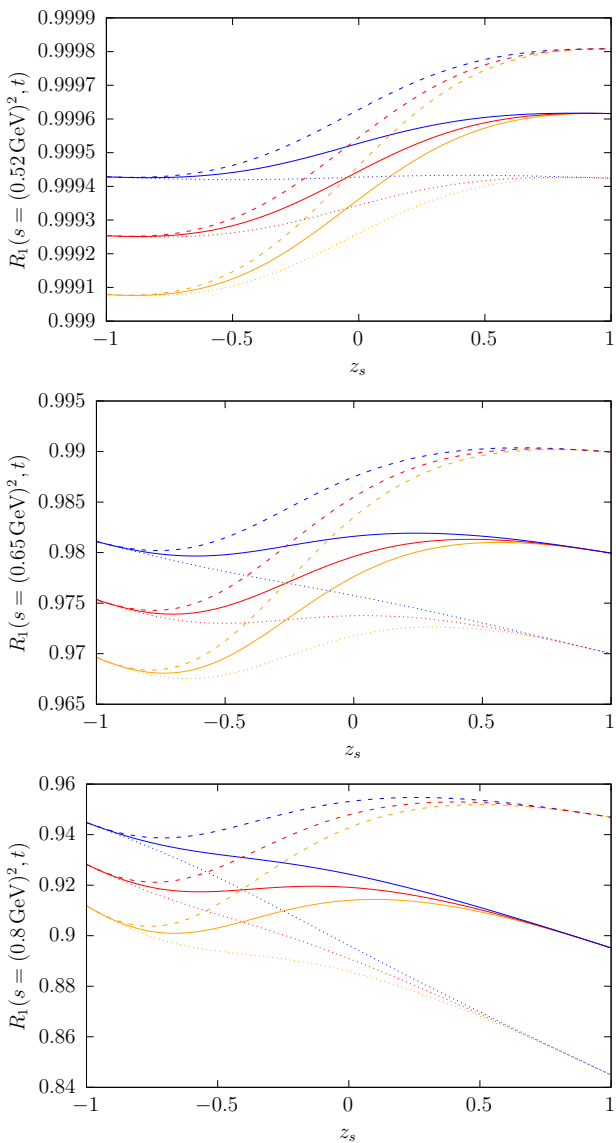

        \centering
        \fontsize{12pt}{14pt} \selectfont
        \scalebox{0.662}{\input{./plots/s052.tex}}
        \scalebox{0.662}{\input{./plots/s065.tex}}
        \scalebox{0.662}{\input{./plots/s08.tex}}
        \caption{Angular dependence of the ratio of differential cross sections $R_1(s,t)$ as defined in Eq.~\eqref{eq:ratio} for three different $s=(0.52\GeV)^2$, $(0.65\GeV)^2$, and $(0.8\GeV)^2$ (top to bottom). As before we employ the ChPT prediction for the polarizabilities and add $0.5$ (orange; dashed), $1$ (red; solid), and $1.5$ (blue; dotted) times the VMD induced corrections.
        Different colors indicate the variation in \mbox{$(\alpha_1-\beta_1)_{K^\pm}$}, whereas different dash-types denote the varied values for \mbox{$(\alpha_1+\beta_1)_{K^\pm}$}. 
        }
        \label{fig:z_variation}
    \end{figure}

AMBER might be able to extend the angular range to forward directions, or positive $z_s$~\cite{Friedrich2024}. In Fig.~\ref{fig:z_variation} we show the $z_s$ dependence of the ratio $R_1(s,t)$, cf.\ Eq.~\eqref{eq:ratio}, for different $s$ and variations of the polarizabilities. 
In accordance with Eq.~\eqref{eq:crosssection_direction},
at backwards angle $z_s=-1$ only $(\alpha_1-\beta_1)_{K^\pm}$ affects the ratio, while at forward angle $z_s=+1$ only $(\alpha_1+\beta_1)_{K^\pm}$ can be extracted. Clearly, good sensitivity at forward angles would enable one to determine $(\alpha_1+\beta_1)_{K^\pm}$ much more precisely, and hence ultimately disentangle electric and magnetic dipole polarizabilities.
Obviously, the same variation of the VMD contribution to $(\beta_1)_{K^\pm}$ has a larger relative impact on $(\alpha_1+\beta_1)_{K^\pm}$ than on $(\alpha_1-\beta_1)_{K^\pm}$. Additionally, the variation in $(\alpha_1-\beta_1)_{K^\pm}$ seems to only have impact for all relevant $s$ starting at $z_s\lesssim 0.5$, while $(\alpha_1+\beta_1)_{K^\pm}$ affects the ratio visibly already for $z_s \gtrsim -0.75 $.

Furthermore, we checked the stability of all these ratios by varying the input $\gamma K\to K\pi$ amplitudes in the dispersion integral within their uncertainty bands. Figs.~\ref{fig:integrate_back}, \ref{fig:zm1}, and \ref{fig:z_variation} do not change at all, since the dispersion integral is included in both the numerator and denominator of the ratios. However, this will obviously be different once the numerator is fixed from experiment. Additionally, Fig.~\ref{fig:integrate_back_nodisp} shows a dependence on the different inputs as the dispersion integral is only used in the numerator.

Note that an experimental extraction of the accompanying pion-production process $\gamma K\to K\pi$ is necessary to properly constrain the amplitudes that are used in the dispersion integral for the resonant $P$-wave. This is also achievable with the AMBER experiment~\cite{Dax:2020dzg}. 
Furthermore, note that for the analysis of the charged-pion polarizabilities~\cite{COMPASS:2014eqi}, also radiative corrections have been taken into account.  These can easily be adapted for charged kaons~\cite{Kaiser:2008jm}, which ought to be done for the experimental analysis.

\section{Neutral-kaon Compton scattering}\label{sec:neutral_kaon}
Compton scattering can also be discussed for \textit{neutral} kaons, and measurements of this kind might become feasible in the future at the planned K-Long Facility (KLF) at Jefferson Lab~\cite{KLF:2020gai}.  Due to the absence of Born terms, the near-threshold amplitude is directly proportional to the dipole polarizabilities, leading to a cross section of the form
    \begin{align}
        \sigma(s)&=\frac{\pi (s-M_K^2)^4}{12M_K^2 s^3}\Big(M_K^4\left(\alpha_1-\beta_1\right)^2_{K^0}\notag\\
        &\hspace{3cm}+s^2\left(\alpha_1+\beta_1\right)^2_{K^0}\Big)\,.\label{eq:cross_section_expansion}
    \end{align}
This would in principle allow for a much more straightforward extraction of these; the downside, obviously, is that the corresponding cross sections are smaller by orders of magnitude.

In ChPT, the first contribution to the neutral-kaon Compton amplitude arises at one-loop order and is given by~\cite{Guerrero:1997rd}
    \begin{align}
        \A^n(s,t)&=- \frac{1}{8\pi^2F_K^2}\Bigg[1-\frac{2M_\pi^2}{t}\arctan^2\bigg(\frac{1}{\sigma^\pi(t)}\bigg)\notag\\
        &\hspace{2cm}-\frac{2M_K^2}{t}\arctan^2\bigg(\frac{1}{\sigma^K(t)}\bigg)\Bigg]\,,\notag\\
        \B^n(s,t)&=0\,.\label{eq:neutral_chpt}
    \end{align}
Both polarizabilities in fact vanish at this order.
We therefore model them, in the spirit of the preceding analysis for charged kaons, by $K^*$-exchange, leading to a nonvanishing magnetic polarizability given by the same expression as in Eq.~\eqref{eq:VMD_pola}. 
With radiative width and masses adjusted to the neutral channel, this results in $(\beta_1)^\text{VMD}_{K^0}=0.8\cdot 10^{-4}\fm^3$. 
The resonant lineshape of the $K^*$ is dispersively reconstructed in direct analogy to the charged case.

The pion loops $\A^\pi(t)$ in Eq.~\eqref{eq:neutral_chpt} are replaced by the dispersive $t$-channel $\gamma\gamma\to\pi\pi\to \bar{K}K$ amplitude following the discussion in Sec.~\ref{sec:Born}. This results in
    \begin{align}
        \A^n(t)&=-\frac{\sqrt{2}}{t}k^0_{++}(t) \\
        &\quad-\frac{1}{8\pi^2F_K^2}\Bigg[\frac{1}{2}-\frac{2M_K^2}{t}\arctan^2\bigg(\frac{1}{\sigma^K(t)}\bigg)\Bigg]\,.\notag
    \end{align}

\begin{figure}[t]
        \centering
        \fontsize{12pt}{14pt} \selectfont
        \scalebox{0.662}{\input{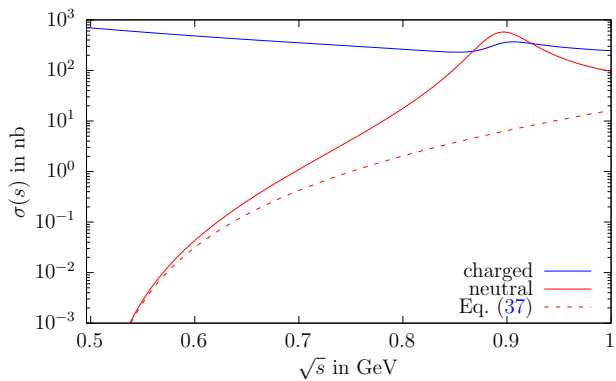}}
        \caption{Cross section for the charged (blue) and neutral (red) kaon Compton scattering. The dashed red line represents Eq.~\eqref{eq:cross_section_expansion}.}
        \label{fig:cross_section}
    \end{figure}

The cross section dominated by the polarizablities in the low-energy region is shown in Fig.~\ref{fig:cross_section}. The solid red line represents the full neutral solution with polarizabilities, loop corrections, and the $K^*$ resonance. The dashed line denotes the result based solely on polarizabilities according to Eq.~\eqref{eq:cross_section_expansion}.
We observe that the polarizabilities dominate up to about $\sqrt{s}=0.7\GeV$; in this range, the total cross section amounts to less than a nanobarn. Compared to the approximately $700\,\text{nb}$ in the charged-kaon case, one looses many orders of magnitude of events. Therefore, experimental observation close to threshold will be very challenging. However, in the region of the $K^*$ resonance the neutral cross section exceeds the charged one due to the larger radiative width of the $K^{*0}$.

\section{Conclusion}\label{sec:conclusions}
We propose a method to extract kaon polarizabilities from differential cross sections of kaon Compton scattering. To this end, we use the ChPT amplitudes up to next-to-leading order as a starting point. Therein, the $t$-channel pion loop contribution that limits the range of applicability of the polarizability expansion can be dispersively improved, employing solutions of a coupled-channel analysis of $\gamma\gamma\to\{\pi\pi,\bar KK\}$. The dominant $K^*$ resonance in the $s$-channel is dispersively reconstructed using amplitudes from the pion photoproduction reaction on kaons, $\gamma K\to K\pi$. 
In contrast to the pion case the hierarchy $(\alpha_1+\beta_1)\ll (\alpha_1-\beta_1)$ is not viable for kaon polarizabilities and both linear combinations need to be considered.
We then suggest a ratio, optimized for sensitivity to the kaon polarizabilities, by incorporating all of the former effects. We show that this ratio enlarges the energy range to be similar to the pion case.  In addition, the relative size of polarizability difference and sum is such in the charged-kaon case that an extension of the experimental analysis of Compton scattering towards forward angles may realistically allow for the disentanglement of $(\alpha_1)_{K^\pm}$ and $(\beta_1)_{K^\pm}$. For neutral kaons the cross section close to threshold is directly proportional to the square of the polarizabilities, however, experimental extraction will become challenging due to the small overall cross section, since there are no corresponding Born terms.

This article provides all necessary theoretical methods for a combined analysis of kaon Primakoff data for both kaon--photon and kaon--pion~\cite{Dax:2020dzg} final states, giving combined access to the chiral anomaly in a kaon reaction, the radiative width of the $K^*(892)$ resonance, and to kaon polarizabilities.  Once experimental data on these reactions is available, e.g., from AMBER, such an analysis ought to be performed to minimize theoretical bias and model dependence. 

\begin{code}
\bsp
Code for the solution of Eq.~\eqref{eq:disp_integral} can be found at \href{https://github.com/HISKP-ph/kaon_polarizabilities}{github.com/HISKP-ph/kaon\_polarizabilities}~\cite{kaon_Compton_HISKP:2024}. Note that there are two versions for the $\gamma K\to K\pi$ amplitudes. First, the solutions for the partial waves of Ref.~\cite{Dax:2020dzg} are stored in \href{https://github.com/HISKP-ph/kaon_polarizabilities/tree/master/gammaKKpi_amp}{/gammaKKpi\_amp}. Secondly, the basis functions are stored in \href{https://github.com/HISKP-ph/kaon_polarizabilities/tree/master/gammaKKpi_amp/basisfunctions}{/gammaKKpi\_amp/basisfunctions}, which allow for a variation of the subtraction constants calculated in Ref.~\cite{Dax:2020dzg}. The code provides a class that allows calculating the first from the second. (This is only true for the mean solution, the uncertainties for the partial waves are only provided via the first version.) Furthermore, the $k^0_{++}$ amplitude for the dispersive $t$-channel calculation is provided in \href{https://github.com/HISKP-ph/kaon_polarizabilities/tree/master/dispersive_t_channel}{/dispersive\_t\_channel}.
\esp
\end{code}

\begin{acknowledgements}
\bsp
We thank Jan Friedrich and Dominik Ecker for helpful discussions about the experimental analysis.
Financial support by the DFG through the funds provided to the Sino--German Collaborative Research Center TRR110 ``Symmetries and the Emergence of Structure in QCD'' (DFG Project-ID 196253076 -- TRR 110) and by the MKW NRW under the funding code NW21-024-A is gratefully acknowledged.

\esp
\end{acknowledgements}

\numberwithin{equation}{section}
\begin{appendix}
\section{Dispersive reconstruction of the \texorpdfstring{$K\pi$}{K pi} intermediate state}\label{app:disc}
We derive the discontinuity equation for a $K\pi$ intermediate state of the kaon Compton scattering amplitude. The definitions of the momenta are shown in Fig.~\ref{fig:dispersive_s-channel}. We start from the unitarity of the $S$-matrix, which leads to 
    \begin{equation}
      \disc(\M_{if})=i\sum_n (2\pi)^4 \delta^{(4)}(q_1+p_1-k_n)\M_{in}\M_{nf}^*\,.
    \end{equation}
The sum in general runs over all possible hadronic intermediate states. Replacing the sum by the appropriate integrations results in
    \begin{align}
        &\disc(\M_{if})=\\
        &\frac{i}{(2\pi)^2}\int \frac{\dd^3 k_1}{2k_1^0}\frac{\dd^3 k_2}{2k_2^0} \delta^{(4)}(q_1+p_1-k_1-k_2)\M_{in}\M_{nf}^*\,.\notag
    \end{align}
The initial-to-final Compton scattering amplitude can be expanded into partial waves via~\cite{Jacob:1959at}
    \begin{align}
        \F_{+\pm}(s,t,u)&\\
        =\sum_{J=1}^\infty(2J&+1)\left(\frac{(s-M_K^2)^2}{4}\right)^Jf_{J,\pm}(s)d_{1,\pm}^J(z_s)\,,\notag
    \end{align}
where the small Wigner $d$-functions are given by
    \begin{equation}
        d_{1,\pm}^J=\frac{1\mp z}{J(J+1)}P'_J(z_s)\pm P_J(z_s)\,.
    \end{equation}
$P_J$ and $P'_J$ are the Legendre polynomials and their derivatives, respectively. The amplitudes involving $K\pi$ intermediate states are of odd intrinsic parity and can be related to scalar amplitudes $\G(s,t,u)$\footnote{For simplicity, we refrain from distinguishing the two possible $K\pi$ charge configurations notation-wise, which will simply be summed over in the final result.}
    \begin{equation}
      \M_{\gamma K \to K \pi,\pm}=i \varepsilon_{\mu\nu\alpha\beta}\epsilon^\mu_{j,\pm}p_i^\nu k_1^\alpha k_2^\beta \G(s,t,u)\,,
    \end{equation}
where $j=1$ for initial to intermediate and $j=2$ for intermediate to final state. The partial-wave expansion of the scalar amplitudes is given by
    \begin{equation}
      \G(s,t,u)=\sum_J g_J(s)P'_J(z_s)\,.
    \end{equation}
$S$-waves are forbidden and the $D$-wave and higher contributions only become relevant outside of the kinematical region we are interested in, where the $K_2^*(1430)$ is the lowest lying resonance starting well beyond $1\GeV$. Therefore, we only consider $P$-waves and the discontinuity relation reads
    \begin{align}
      3\frac{(s-M_K^2)^2}{4}\frac{1\pm z_s}{2}\disc(f_{1,\pm}(s))&\notag\\
        =\frac{i}{(2\pi)^2}\frac{\lambda^{1/2}(s,M_\pi^2,M_K^2)}{8s}& |g_1(s)|^2\mathcal{I}_\pm\,,
    \end{align}
where
    \begin{align}
        \mathcal{I}_\pm=\int \dd \cos(z'_s) \dd \phi' \varepsilon_{\mu\nu\alpha\beta}&\varepsilon_{\sigma\rho\gamma\delta}\epsilon_{1,+}^\mu \epsilon_{2,\pm}^\sigma p_1^\nu p_2^\rho k_1^\alpha k_1^\gamma k_2^\beta k_2^\delta\,. 
    \end{align}
Since $\mathcal{I}_\pm$ is fully contracted, we can evaluate it in the center-of-mass system and choose an explicit representation for the momenta~\cite{Dammann2023}. This results in
    \begin{equation}
        \mathcal{I}_\pm=\pm\pi\frac{1\pm z_s}{24s}\lambda(s,M_\pi^2,M_K^2)(s-M_K^2)^2\,.
    \end{equation}
Therefore the discontinuity of the $P$-wave for the $+\pm$ helicities reads
    \begin{align}
        2i\Im(f_{1,\pm}(s))&=\disc(f_{1,\pm}(s))\notag\\
        =\pm\frac{i}{4\pi}&\frac{\lambda^{3/2}(s,M_\pi^2,M_K^2)}{72s^2}|g_1(s)|^2\,.
    \end{align} 
\end{appendix}

\bibliographystyle{utphysmod}
\bibliography{Literature}

\end{document}